\documentclass[aps,prd,onecolumn,groupedaddress,nofootinbib]{revtex4}         
\usepackage{tabularray}
\usepackage[english]{babel}
\usepackage{booktabs}
\usepackage[letterpaper,top=2cm,bottom=2cm,left=3cm,right=3cm,marginparwidth=1.75cm]{geometry}

\usepackage{amsmath}
\usepackage{graphicx}
\usepackage[colorlinks=true, allcolors=blue]{hyperref}
\usepackage{xcolor}
\usepackage{multirow}

\begin{document}

\title{Constraining the Potential Index $n$ of the Early Dark Energy Model Using Cosmic Birefringence from Planck and ACT}

\author{Kedi Zhang$^{1}$}
\author{Lu Yin$^{1}$}
\email{yinlu@shu.edu.cn}

\affiliation{$^{1}$Department of Physics, Shanghai University, Shanghai, 200444,  China.}

\date{\today}

\begin{abstract}
Cosmic birefringence and the Hubble tension represent compelling challenges to the standard $\Lambda$CDM model. The early dark energy (EDE) model with potentials $V(\phi) \propto [1-\cos(\phi/f)]^n$ offer a unified framework to address both anomalies through energy injection near matter-radiation equality and parity-violating Chern--Simons coupling to photons. While previous studies have focused on $n=3$, the dependence of the birefringence signal on the potential index $n$ remains largely unexplored. We perform a comprehensive statistical analysis of axion-like EDE models with $n=2$, $n=3$, and $n=\infty$, using $EB$ cross-polarization data from Planck-$EB$ and ACT-$EB$. 
The $n=2$ model is severely disadvantaged, displaying extreme coupling values ($gM_{\rm pl} \approx 69.912$ for Planck, $-40.726$ for ACT), large $\chi^2_{\rm min}$ (144.52 and 86.93), and $\Delta \chi^2<1$ with many local minials. Conversely, $n=3$ achieves the best fits ($\chi^2_{\rm min} = 65.70$ and $48.08$) with consistent couplings ($gM_{\rm pl} = -0.210 \pm 0.024$ and $-0.158 \pm 0.025$) that accurately reproduce observations across all angular scales.
  We checked that the $n=3$ configuration represents the optimal choice for simultaneously addressing the Hubble tension and cosmic birefringence within a theoretically viable framework.
\end{abstract}

\maketitle

\section{Introduction}

The standard $\Lambda$CDM cosmological model has achieved remarkable empirical success over the past decades \cite{Abdalla2022,Bernal2016}. However, recent high-precision observations have revealed persistent anomalies that challenge its theoretical completeness \cite{DiValentino2021,Kamionkowski2023,Schoeneberg2022}. Among these, two observational tensions stand out as particularly compelling: the Hubble tension and cosmic birefringence. These phenomena have motivated extensive theoretical work exploring physics beyond the standard framework \cite{Krishnan2021,Luongo2022,LiTN2024,Feng2025,Krishnan2022}.
The Hubble tension refers to the significant ($4$--$5\sigma$) discrepancy between direct late-time measurements of the Hubble constant $H_0$, and the value inferred from cosmic microwave background (CMB) observations under the assumption of $\Lambda$CDM \cite{Riess2022,Escamilla2025,Du2025,Simon2025}. This tension persists across independent measurement techniques and datasets, suggesting potential new physics in the early or intermediate universe \cite{Colgain2021,DAmico2021,Alexander2019,Chudaykin2020}.
In parallel, mounting evidence for cosmic birefringence has emerged from CMB polarization analyses. Recent studies report a non-zero birefringence angle $\beta = 0.342^\circ \pm 0.091^\circ$ at $3.6\sigma$ significance from  Planck satellite data \cite{Eskilt2022,Sullivan2025}. Observationally, cosmic birefringence manifests as a non-zero cross-correlation between the $E$-mode and $B$-mode polarization patterns in the CMB \cite{Feng2005,Li2007,Feng2006,Li2008}. Since the $EB$ cross-power spectrum $C_\ell^{EB}$ is parity-odd, its detection provides direct evidence for parity-violating physics in the early universe \cite{Lue1999,Xia2008a,Xia2008b}. Theoretical frameworks incorporating such parity violation have been extensively developed and confronted with observations \cite{Li2015,Li2013,Zhao2015,Marsh2016,Naokawa2023,Namikawa2023,Greco2024,Ferreira2024}.

The early dark energy (EDE) model represent a promising avenue for simultaneously addressing both the Hubble tension and cosmic birefringence \cite{Karwal2016,Murgia2021,Poulin2019,Efstathiou2024}. The core mechanism involves injecting additional energy density around the epoch of matter-radiation equality, which increases the Hubble expansion rate during this period and consequently reduces the comoving sound horizon at photon decoupling. To preserve the observed angular scale of acoustic peaks in the CMB, this reduction necessitates a larger present-day Hubble constant, thereby alleviating the tension with local measurements \cite{Piras2025,LiTN2026,Akarsu2023,Niedermann2021}. This dynamical framework has been validated through numerical simulations and perturbative analyses across various implementations \cite{Freese2005,Ye2020,Akarsu2020,Lin2019,Braglia2020,Hill2020,HeroldThesis}.
The connection between EDE and cosmic birefringence arises naturally when the underlying scalar field is an axion-like particle coupled to electromagnetism via a Chern--Simons interaction $ g\phi F_{\mu\nu}\tilde{F}^{\mu\nu}$ \cite{Fedderke2019,Choi2021,Nakatsuka2022,Carroll1990,Harari1992}. This parity-violating term induces a differential propagation speed for left- and right-handed photon polarizations, causing a net rotation of the CMB polarization plane as photons traverse the evolving scalar field background. The accumulated rotation generates the observed $EB$ correlation, linking EDE dynamics directly to the birefringence signal \cite{Xia2008a}.

To ensure compatibility with late-time cosmological observations, the EDE scalar field must dilute sufficiently rapidly after matter-radiation equality \cite{Yin2023,Lonappan2025,Lee2025,Yin2025,Kochappan2025,Yin2022,Lin2025}. While existing studies of cosmic birefringence within axion-like EDE frameworks have predominantly adopted $n=3$, the dependence of the birefringence signal on the potential index $n$ remains largely unexplored. A systematic investigation across different values of $n$ is essential for understanding which configurations are favored by observational data and for assessing the robustness of EDE as a unified solution to multiple cosmological tensions.
In this work, we perform an analysis of the EDE model with potential indices $n=2$, $n=3$, and $n=\infty$ using $EB$ cross-polarization measurements from Planck 2018 \cite{Komatsu2014,Aghanim2020,Minami2020} and the Atacama Cosmology Telescope Data Release 6 (ACT DR6) \cite{Madhavacheril2024,Louis2025ACT,Qu2024}. ACT DR6 leverages ground-based observations with an aperture approximately five times larger than Planck, achieving superior angular resolution and extending measurements to higher multipoles of the sky \cite{DiegoPalazuelos2022,DiegoPalazuelos2025,Dutcher2021,Ade2021BICEP,Ade2022SPIDER}. Together, these datasets enable stringent tests of theoretical models across a wide dynamic range.

The remainder of this paper is organized as follows. Section~\ref{sec:theory} reviews the theoretical framework connecting axion-like EDE to cosmic birefringence. Section~\ref{sec:constraints} presents our observational constraints on $gM_{\rm pl}$ for each potential index. Section~\ref{sec:conclusion} summarizes our findings and discusses their implications.

\section{Cosmic Birefringence Induced by Early Dark Energy}
\label{sec:theory}
 { 

Cosmic birefringence can naturally arise from parity-violating interactions between photons and a dynamical pseudoscalar field. In a broad class of axion-inspired models, such interactions are described by a Chern--Simons-type coupling between the scalar field $\phi$ and the electromagnetic field tensor. The effective Lagrangian takes the form \cite{Caldwell2003, Smith2020, Berghaus2020}
\begin{equation}
\mathcal{L} = -\frac{1}{2}(\partial_{\mu}\phi)^{2} - V(\phi) - \frac{1}{4}F_{\mu\nu}F^{\mu\nu} - \frac{1}{4}g\phi F_{\mu\nu}\tilde{F}^{\mu\nu},
\end{equation}
where $V(\phi)$ denotes the scalar-field potential, $F_{\mu\nu}$ is the electromagnetic field strength tensor, and $\tilde{F}^{\mu\nu} = \frac{1}{2}\epsilon^{\mu\nu\alpha\beta}F_{\alpha\beta}$ represents its dual. The coupling constant $g$ determines the interaction strength between photons and the pseudoscalar background. Such parity-violating operators emerge naturally in axion and string-inspired frameworks and provide a well-motivated mechanism for generating cosmological birefringence.

A direct consequence of the Chern--Simons interaction is the breaking of the degeneracy between left- and right-handed photon helicities. In a homogeneous cosmological background, the modified photon dispersion relation becomes \cite{Carroll1991, Schiappacasse2018}
\begin{equation}
\omega_{\pm} = k \mp \frac{g}{2}\dot{\phi},
\end{equation}
where $k$ is the photon wavenumber and $\dot{\phi}\equiv d\phi/dt$ describes the background evolution of the scalar field. The opposite signs associated with the two helicity states imply that photons acquire different propagation phases during their evolution. Consequently, linearly polarized radiation experiences a continuous rotation of its polarization plane, giving rise to the phenomenon of cosmic birefringence.

The rotation angle equals the integral of the helicity-dependent phase difference along the photon trajectory.
Under the WKB approximation, the polarization rotation angle from emission time $t$ to the present epoch $t_0$ is given by \cite{Finelli2009, Galaverni:2014gca}
\begin{equation}
\beta(t) = -\frac{1}{2}\int_{t}^{t_{0}}dt'(\omega_{+} - \omega_{-}) = \frac{g}{2}[\phi(t_0) - \phi(t)],
\end{equation}
shows that the total rotation angle depends only on the difference between the scalar-field values at the endpoints of photon propagation. Thus, CMB photons record the cosmological evolution of $\phi$, making them sensitive probes of parity-violating physics in the early Universe.

To describe the dynamics of the pseudoscalar field, we consider an axion-like EDE scenario motivated by string-inspired constructions. The scalar potential is parametrized as \cite{Kochappan2025, Lin2025}
\begin{equation}
V(\phi) = V_0\left[1 - \cos\left(\frac{\phi}{f}\right)\right]^n,
\end{equation}
where $V_0$ characterizes the energy scale of the potential, and the potential index $n$ can be 2, 3, ..., and $\infty$. \cite{Takahashi2021, Murai2023}. 

The polarization of CMB radiation is conventionally described by the Stokes parameters. In standard cosmology,  the polarization state is characterized by the intensity $I$ and the linear polarization parameters $Q$ and $U$, which can be written as \cite{Xia2008a, Xia2010, Xia2008b}

\begin{equation}
I_{ij} =
\begin{pmatrix}
I+Q & U \\
U & I-Q
\end{pmatrix},
\end{equation}
where $Q$ and $U$ describe the magnitude and orientation of linear polarization.

A time-dependent birefringence angle modifies the Boltzmann equation of CMB polarization. In conformal time $\eta$, the spin-$\pm2$ polarization perturbations ${}_{\pm2}\Delta_P$ satisfy \cite{Lesgourgues2011, Blas2011}
\begin{equation}
{}_{\pm 2}\Delta_P' + i q \mu {}_{\pm 2}\Delta_P = \tau' \left[ - {}_{\pm 2}\Delta_P + \sqrt{\frac{6\pi}{5}} {}_{\pm 2}Y_2^0(\mu)\Pi(\eta, q) \right] \pm 2i\beta' {}_{\pm 2}\Delta_P,
\end{equation}
where a prime denotes differentiation with respect to conformal time, $\tau'$ is the differential optical depth, and $\Pi$ is the Thomson-scattering source term. Compared with the standard case, the additional term proportional to $\beta' \equiv d\beta/d\eta$ rotates linear polarization, mixing $E$ and $B $ modes and generating nonzero $TB$ and $EB$ correlations.

In the limit where the birefringence angle varies sufficiently slowly after recombination, the cumulative rotation effect can be approximated by an effective angle $\beta$. In this case, the observed $E$- and $B$-mode polarization are mixed, giving rise to a nonzero $EB$ correlation. The corresponding angular power spectrum can be written as \cite{Li2015, Li2013}
\begin{equation}
C_{\ell}^{EB} = \frac{1}{2}\sin(4\beta)(\tilde{C}_{\ell}^{EE} - \tilde{C}_{\ell}^{BB}).
\end{equation}
In the standard $\Lambda$CDM model, parity conservation implies $C_{\ell}^{EB}=0$. Therefore, a nonzero $EB$ correlation may indicate parity-violating effects in the photon propagation.
For comparison with CMB observations, it is convenient to introduce the dimensionless angular power spectrum $D_{\ell}^{EB}$ as
\begin{equation}
D_{\ell}^{EB} = \frac{\ell(\ell+1)}{2\pi} C_{\ell}^{EB}.
\end{equation}
We will discuss the comparison between theoretical power spectrum values and experimental data (Planck-$EB$ and ACT-$EB$) in the next section.}

\section{Constraints on Chern-Simons Coupling Constant with Observations}
\label{sec:constraints}

\begin{table}[h!]
\centering
\renewcommand{\arraystretch}{1.3} 
\caption{Best-fit results of $gM_{pl}$ and minimum $\chi^2$ results for different potential indices $n$ under the Planck-$EB$ and ACT-$EB$ datasets.}
\label{tab:bestfit}  
\begin{tabular}{|c|c|c|c|}
\hline
\textbf{Dataset} & \textbf{$n$} & \textbf{$gM_{pl}$} & \textbf{$\chi^2$} \\
\hline
\multirow{3}{*}{Planck-$EB$} & $n=2$ & $69.912$ & $144.52$ \\
\cline{2-4}
            & $n=3$ & $-0.210 \pm 0.024$ & $65.70$ \\
\cline{2-4}
            & $n=\infty$ & $-0.017 \pm 0.002$ & $68.38$ \\
\hline
\multirow{3}{*}{ACT-$EB$} & $n=2$ & $-40.726$ & $86.93$ \\
\cline{2-4}
        & $n=3$ & $-0.158 \pm 0.025$ & $48.08$ \\
\cline{2-4}
        & $n=\infty$ & $-0.012 \pm 0.002$ & $49.67$ \\
\hline
\end{tabular}
\end{table}

\begin{figure}[t]
\centering
\begin{tabular}{ccc}
\includegraphics[width=0.5\linewidth]{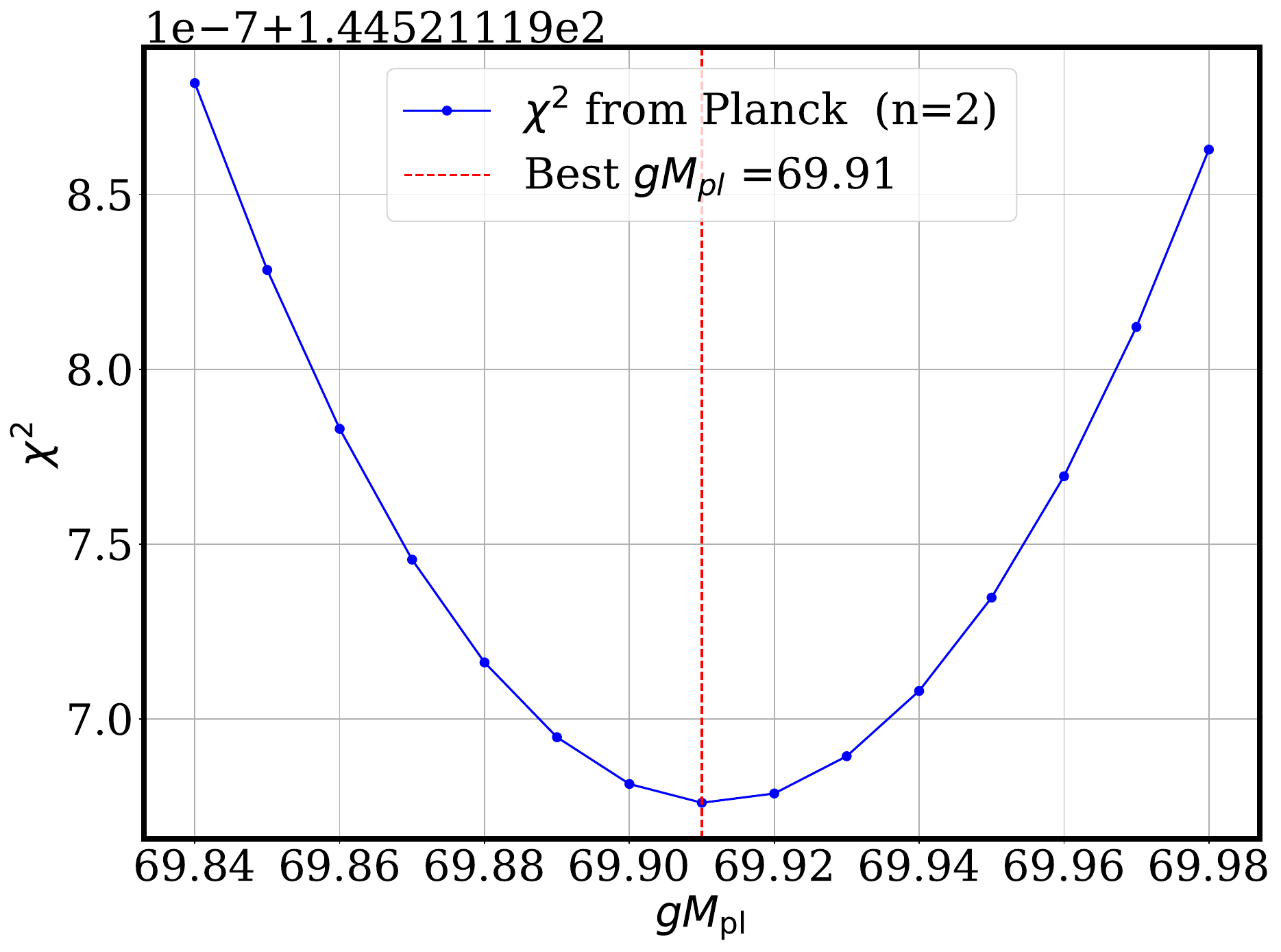} &
\includegraphics[width=0.49\linewidth]{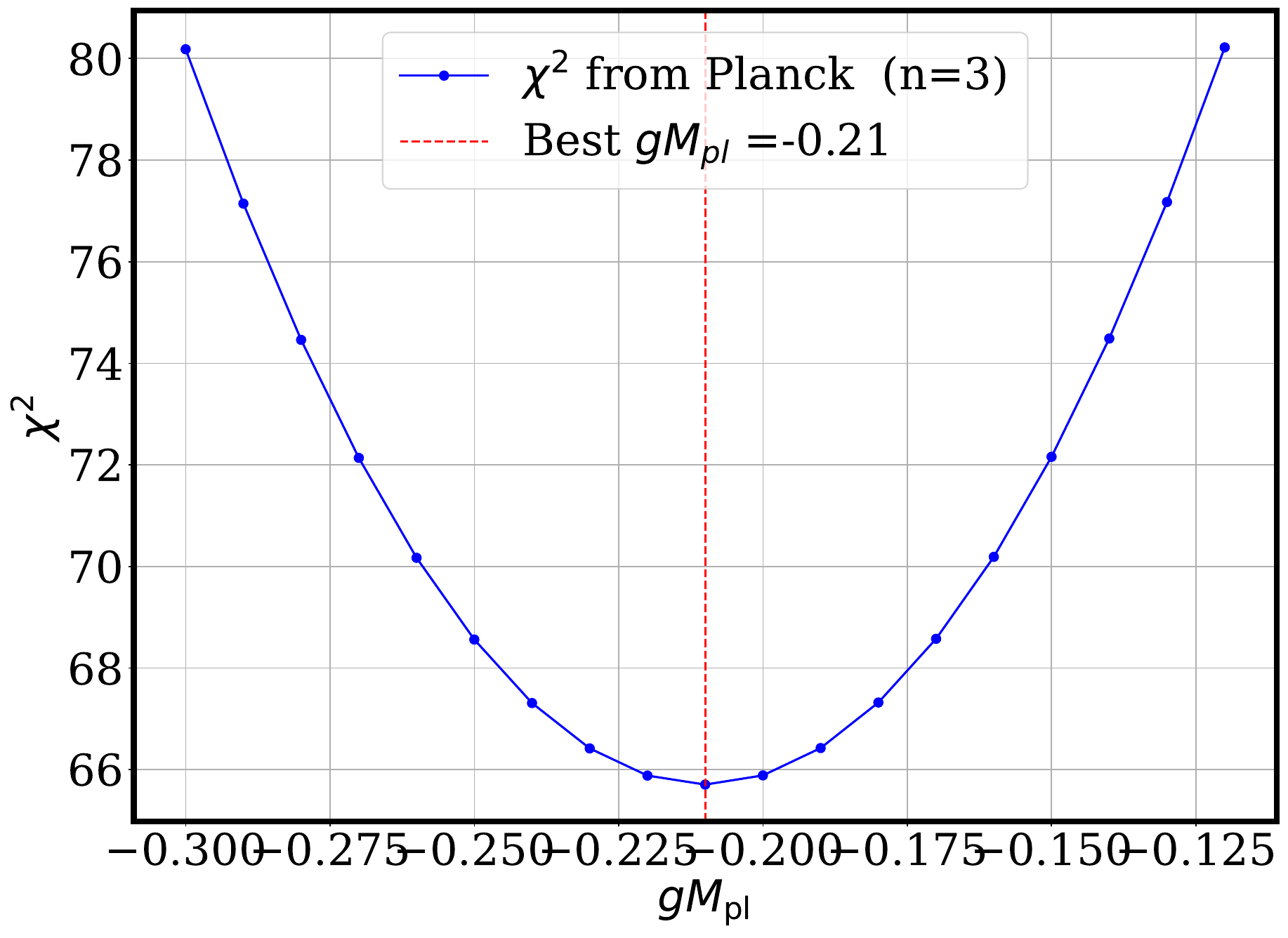} \\
(a)  $n=2$ & (b) $n=3$ \\
\includegraphics[width=0.52\linewidth]{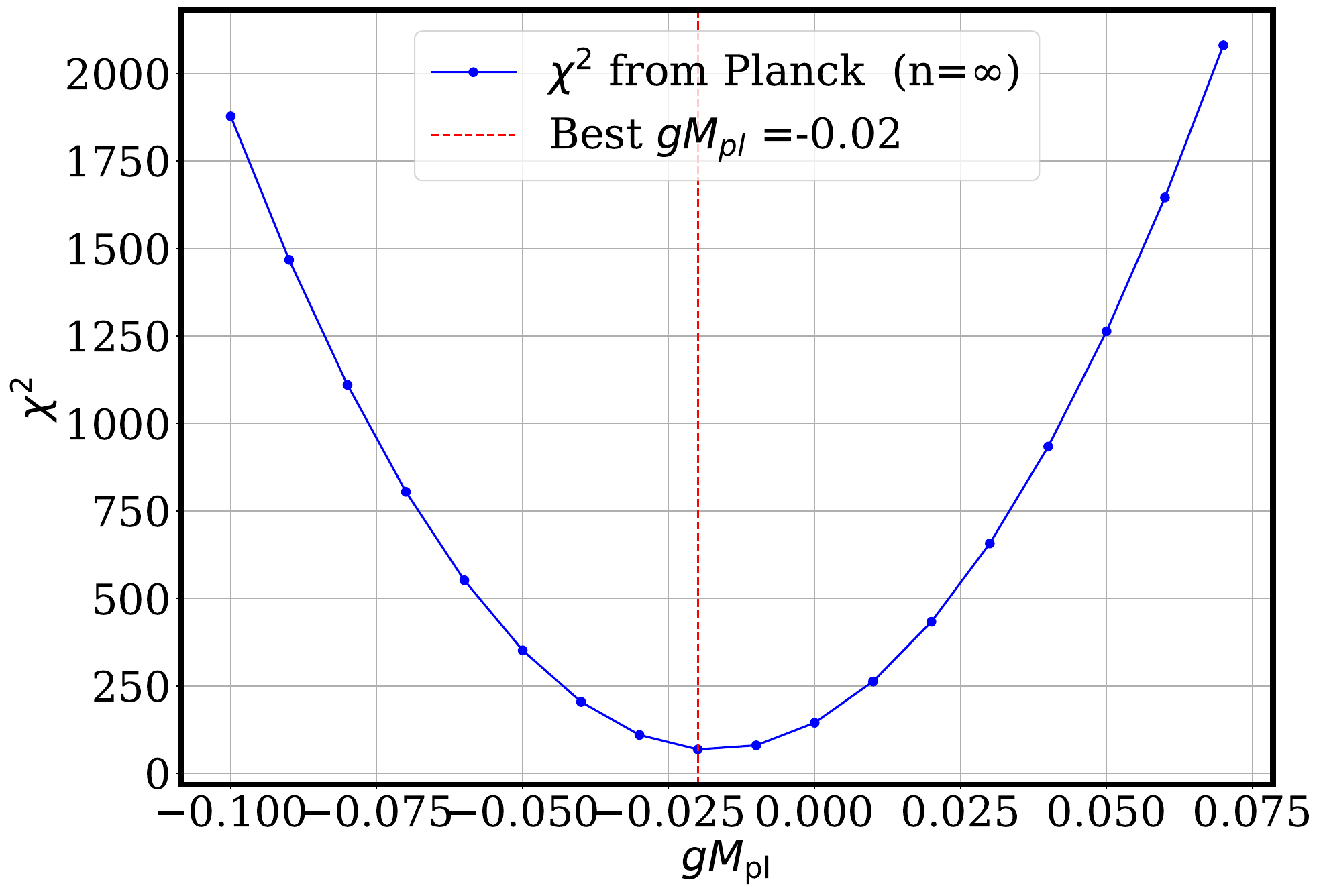}\\
(c) $n=\infty$
\end{tabular}
\caption{\label{fig:planck_chi2}Distribution curves of $\chi^2$ versus the Chern-Simons coupling constant $gM_{pl}$ for the EDE model with Planck-$EB$ data. (a), (b) and (c) are the EDE model with potential index $n=2$, $3$, $\infty$, respectively.}
\end{figure}

\begin{figure}[t]
\centering
\begin{tabular}{ccc}
\includegraphics[width=0.5\linewidth]{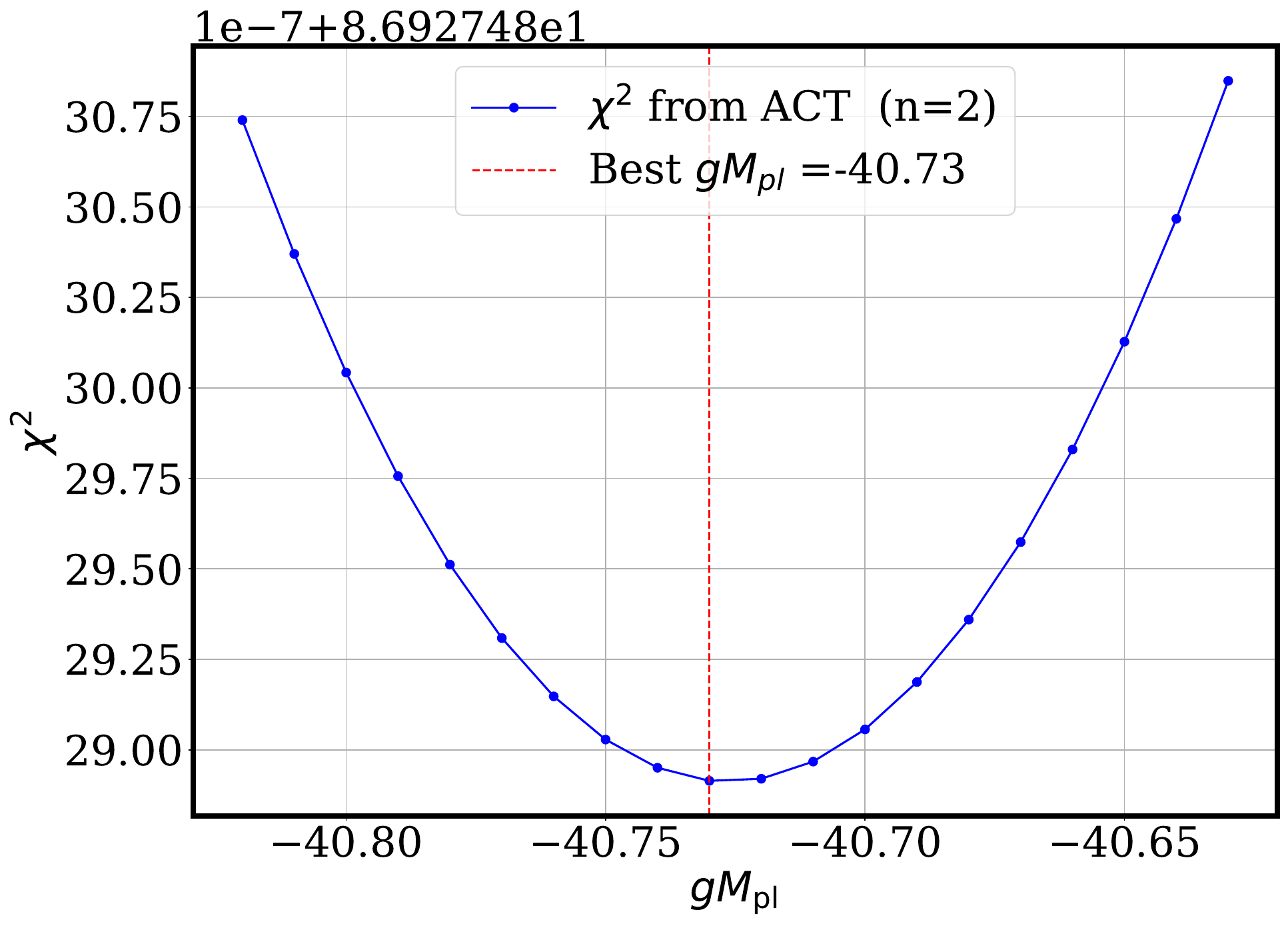} &
\includegraphics[width=0.49\linewidth]{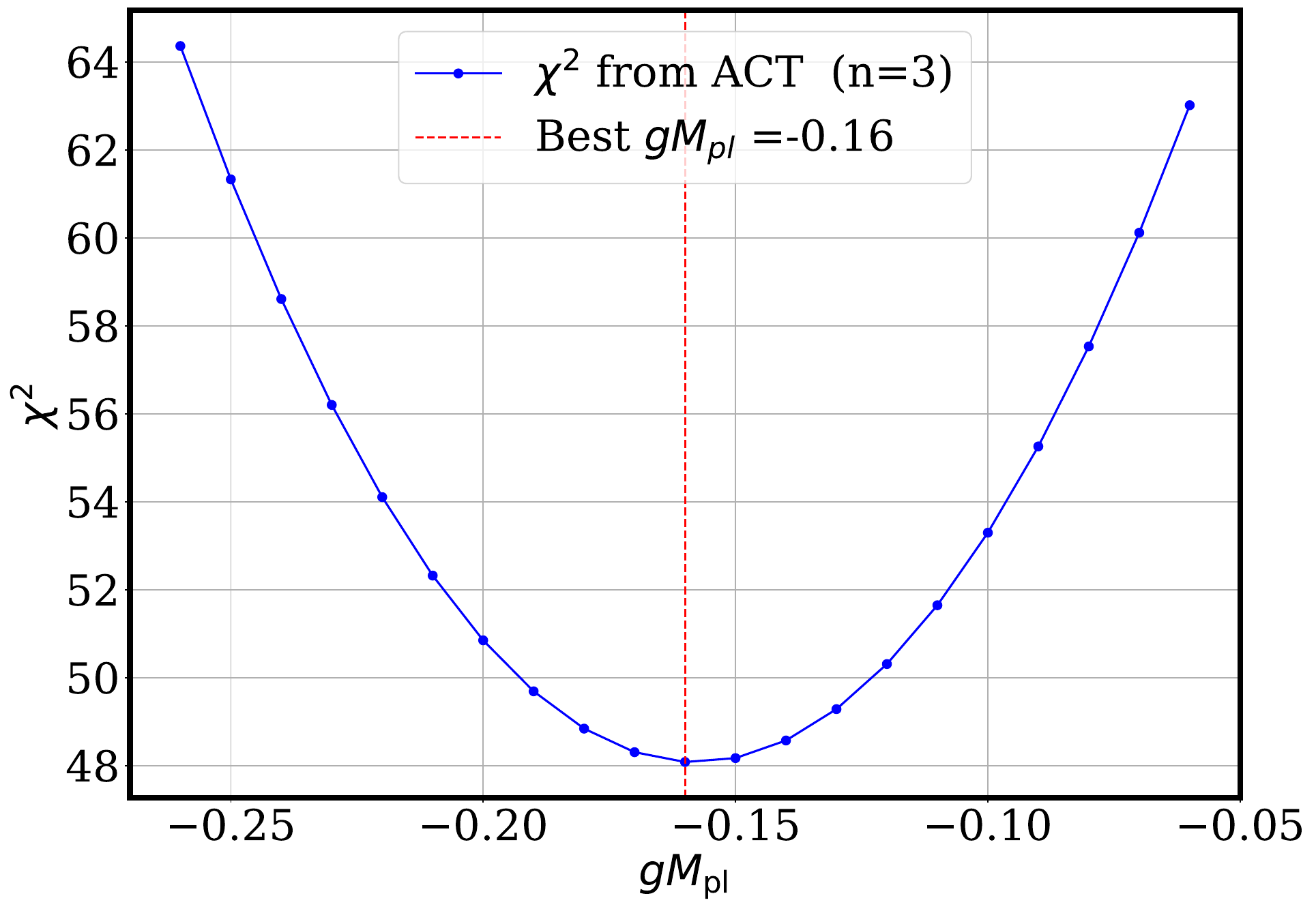} \\
(a)  $n=2$ & (b) $n=3$ \\
\includegraphics[width=0.52\linewidth]{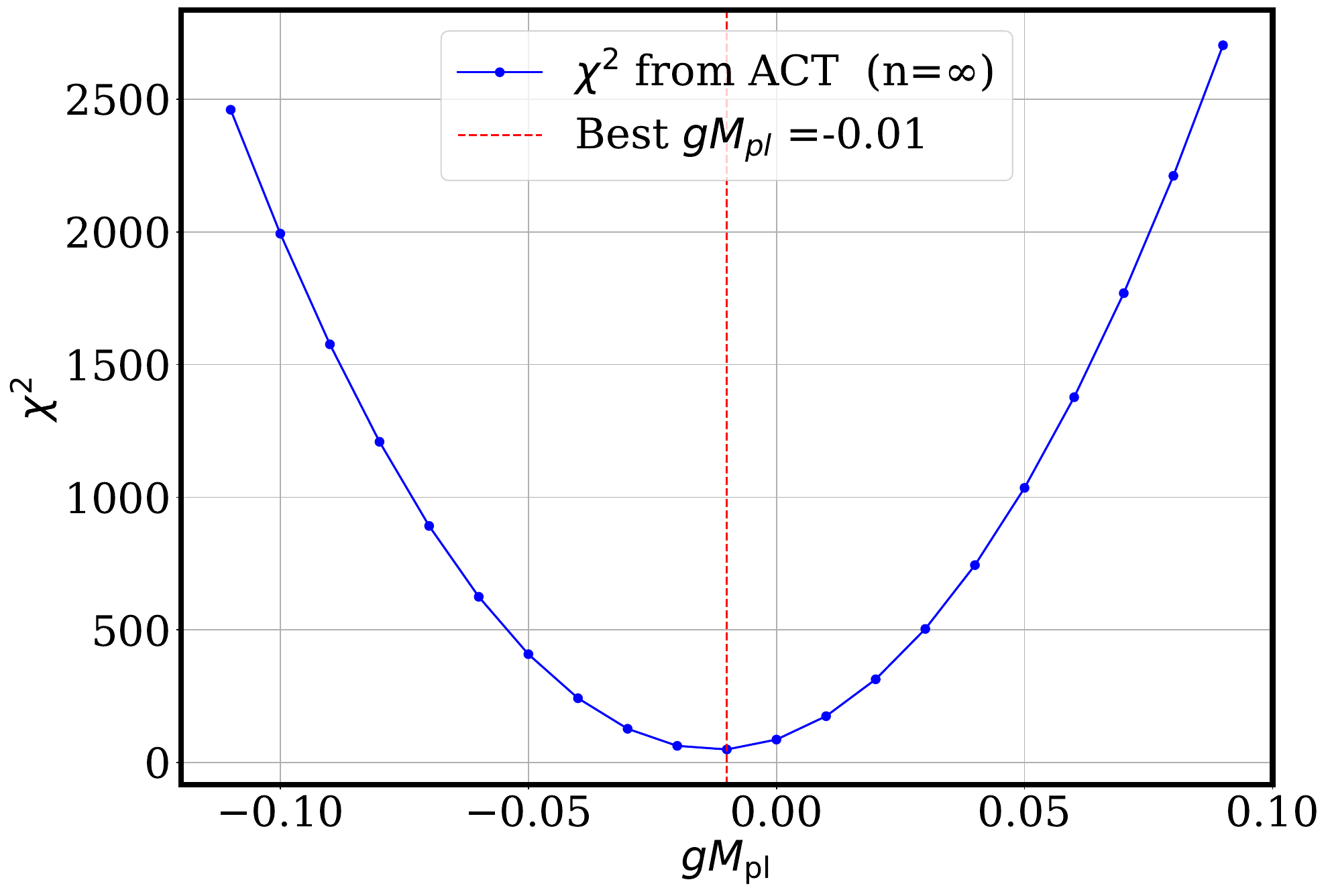}\\
(c) $n=\infty$
\end{tabular}
\caption{\label{fig:act_chi2}Distribution curves of $\chi^2$ versus the coupling parameter $gM_{pl}$ for the EDE model under different potential indices ($n=2, 3, \infty$), utilizing ACT-$EB$ data.}
\end{figure}

\begin{figure}[t]
    \centering
    \begin{tabular}{cc}
    \includegraphics[width=0.49\textwidth]{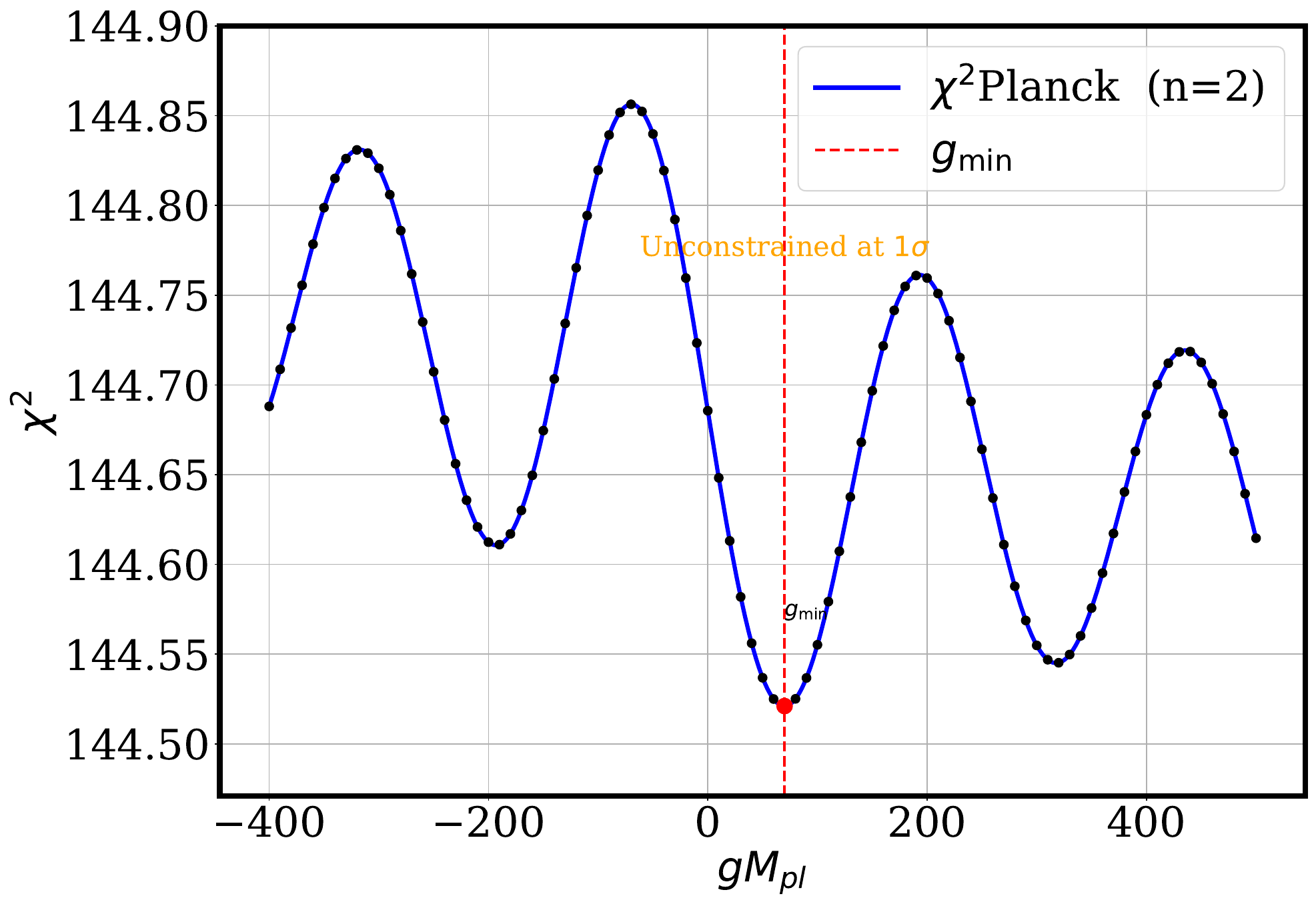} &
        \includegraphics[width=0.49\textwidth]{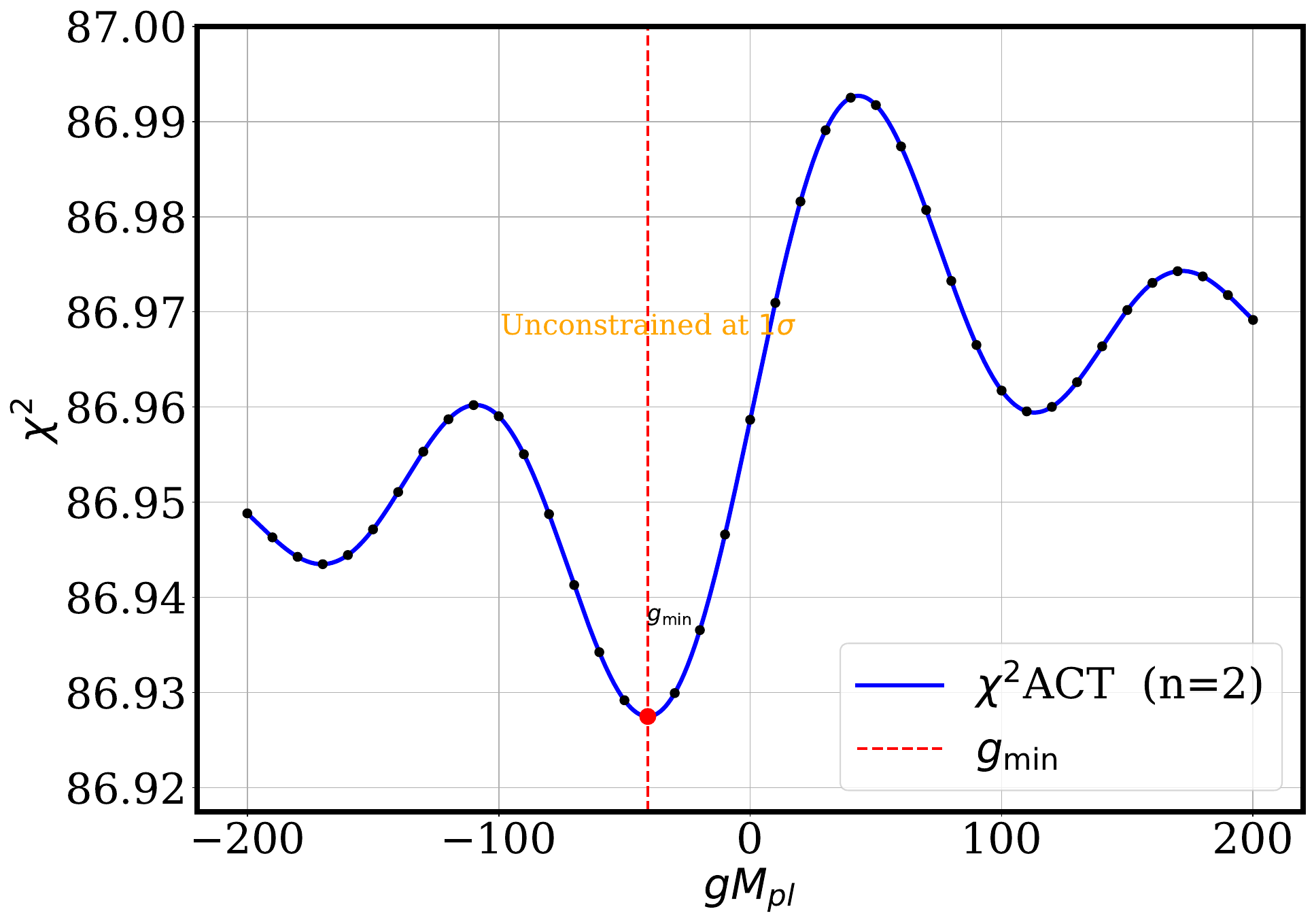} \\
        (a)  $\chi^2$ of EDE $n=2$  from Planck-$EB$, & (b)  $\chi^2$ of EDE $n=2$  from ACT-$EB$. \\
    \end{tabular}
    \caption{The $\chi^2$ profile for the coupling parameter $gM_{\mathrm{pl}}$, constrained by the Planck $EB$ (a) and ACT-$EB$ (b) polarization power spectrum, assuming an EDE model with $n=2$.}
    \label{fig:chi2_gMpl_Planck_n2}
\end{figure}

\begin{figure}[t]
    \centering
    \includegraphics[width=0.65\textwidth]{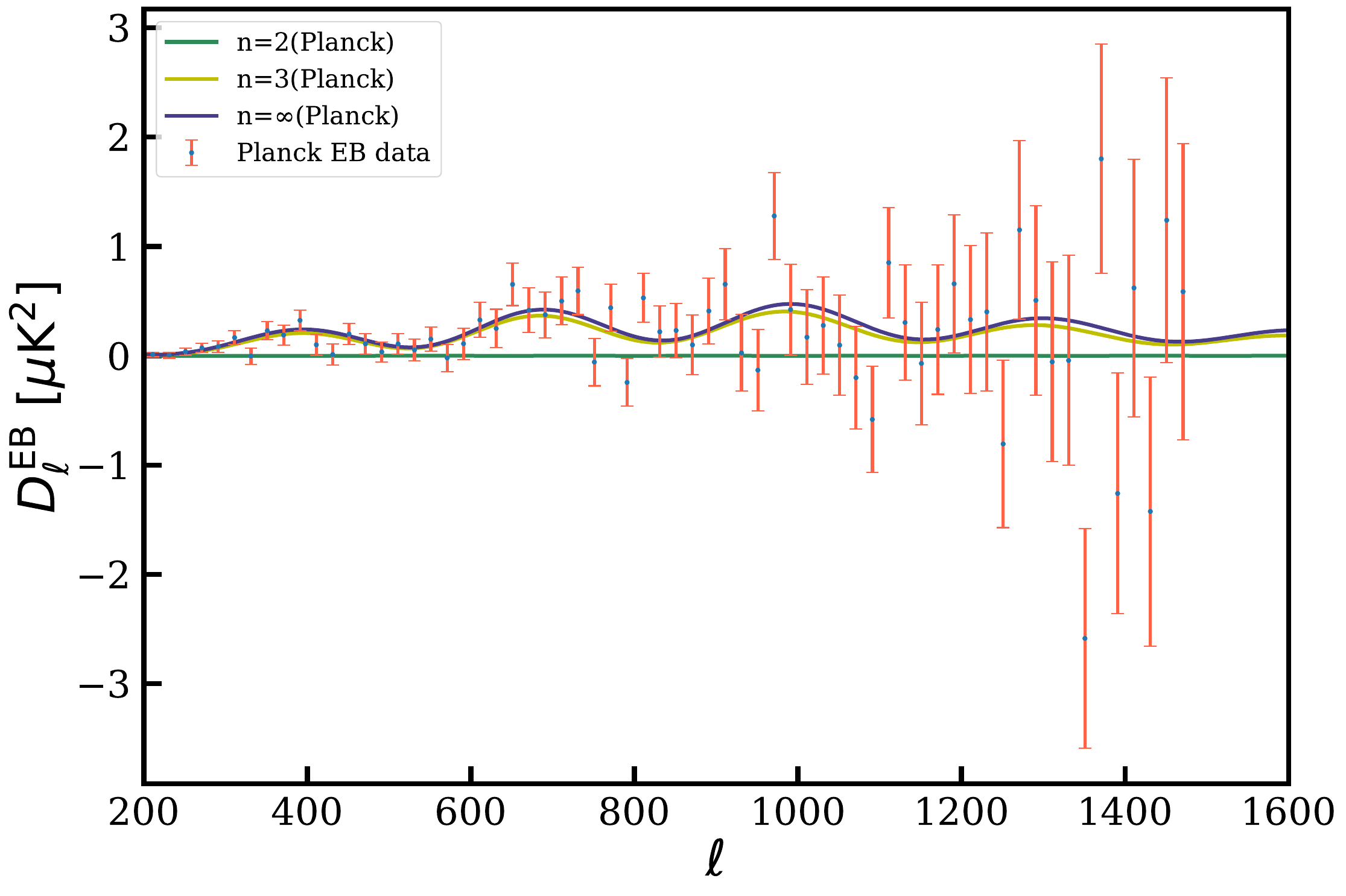} 
    \caption{\label{fig:planck_spectrum} CMB $EB$ power spectrum  $D_\ell^{EB}$ compare with Planck-$EB$ data.}
\end{figure}    

\begin{figure}[t]
    \centering 
    \includegraphics[width=0.65\textwidth]{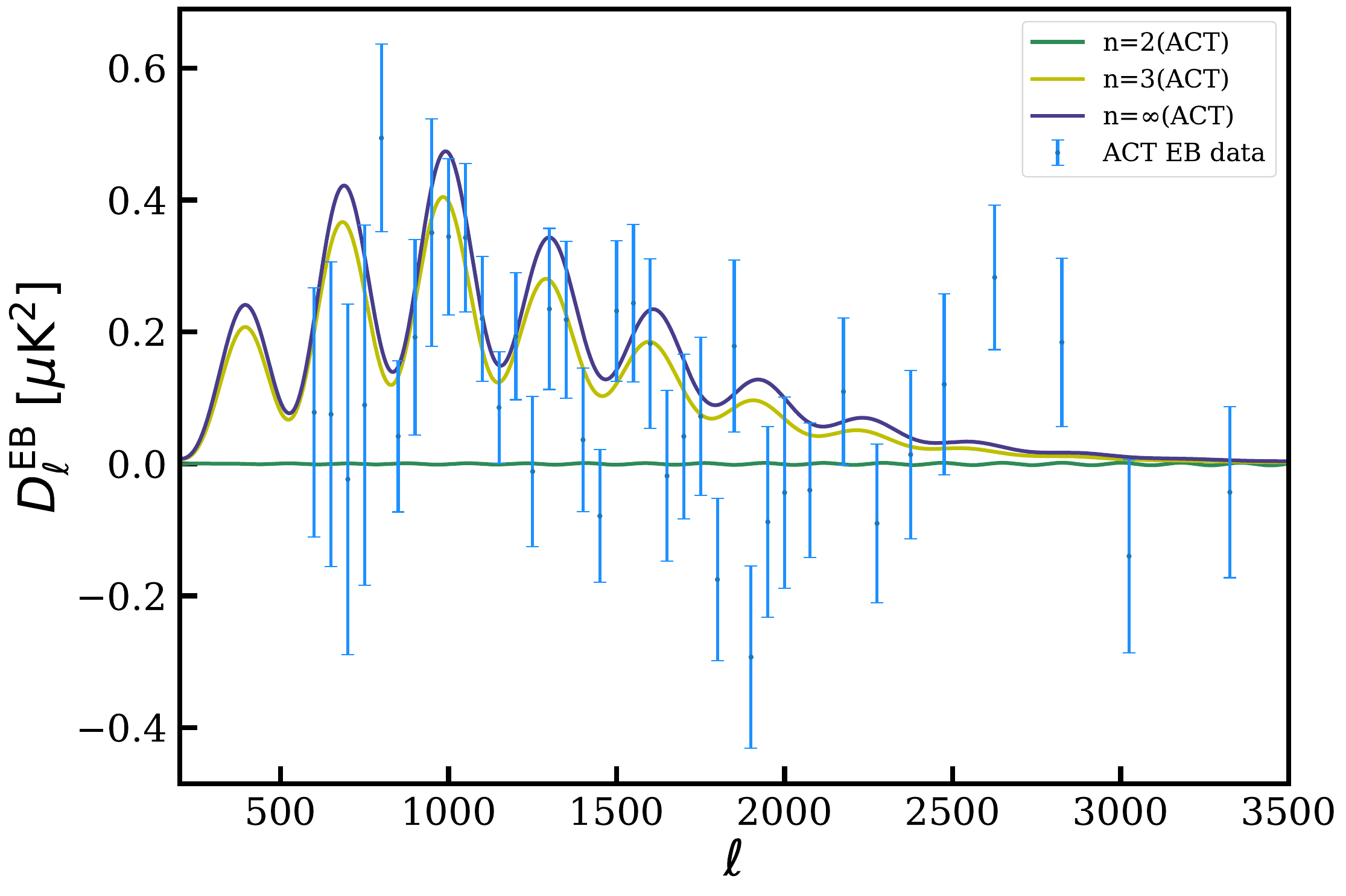}
    \caption{\label{fig:morphology_spectra}CMB $EB$ power spectrum  $D_\ell^{EB}$ compare with  ACT-$EB$ data.}
\end{figure}

\begin{figure}[t]
    \centering
    \includegraphics[width=0.65\textwidth]{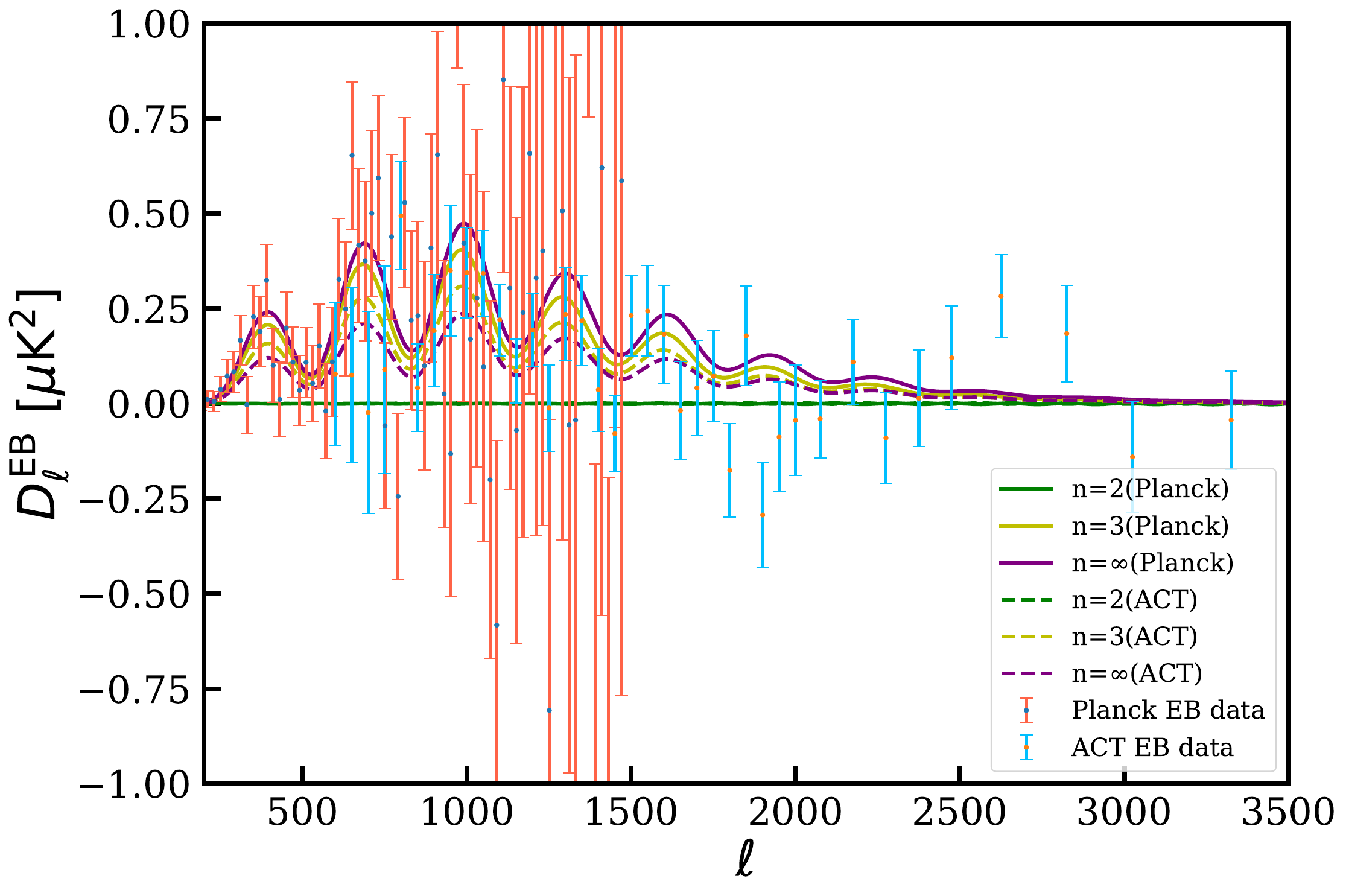}
    \caption{\label{fig:combined_spectrum}Combined morphological comparison of theoretical $D_\ell^{EB}$ power spectra for different EDE potential indices against Planck-$EB$ (red) and ACT-$EB$ (light-blue) data. Solid and dashed curves represent models' $gM_{pl}$ constrained by Planck and ACT respectively. Colors indicate potential indices: green ($n=2$), yellow ($n=3$), and purple ($n=\infty$). }
\end{figure}

 \begin{figure}[t]
    \centering
    \includegraphics[width=0.65\textwidth]{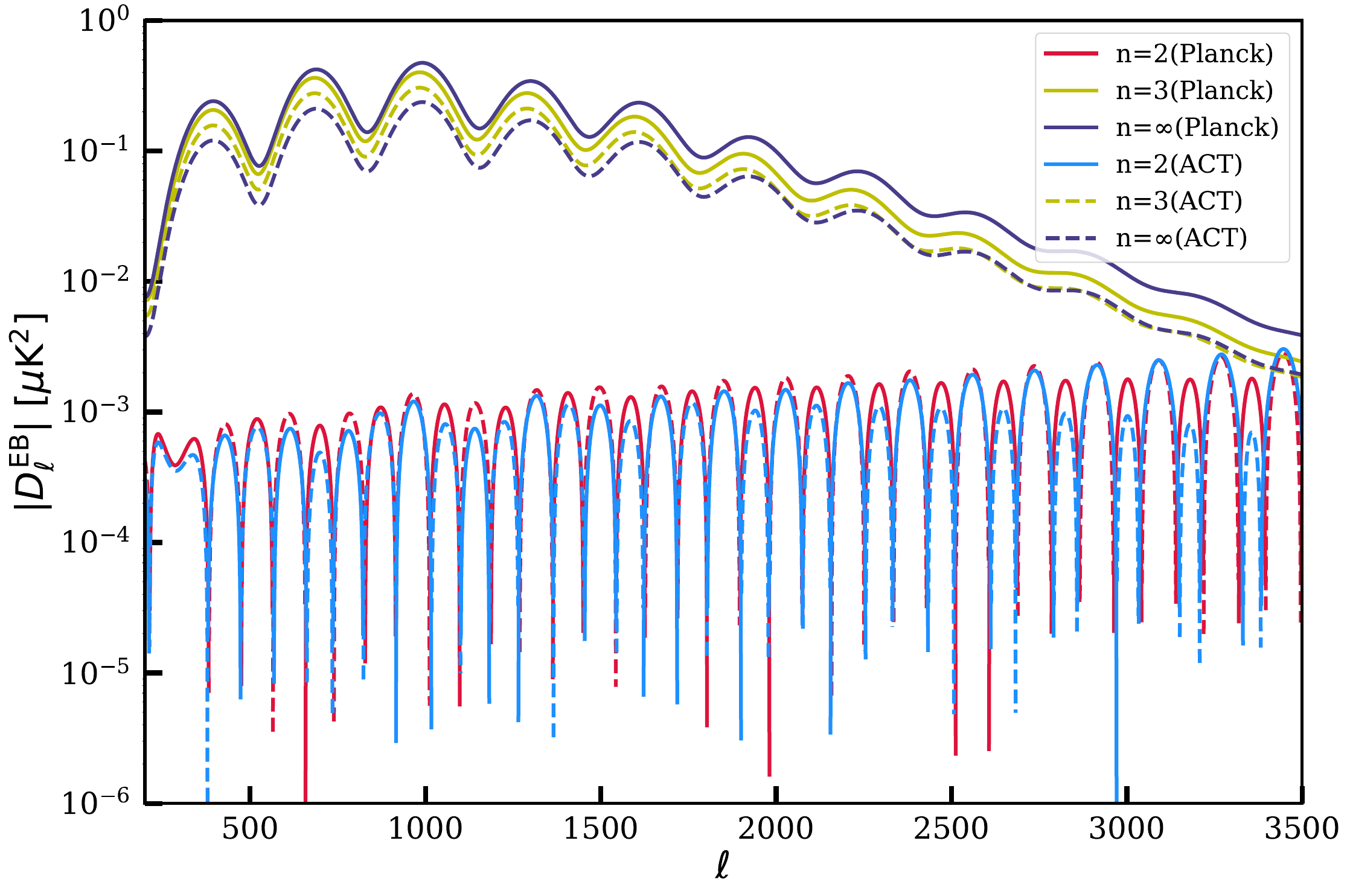}
    \caption{\label{fig:log_spectrum}Logarithmic absolute spectrum ($|D_\ell^{EB}|$) from EDE with $n=2$, 3, $\infty$, respectively. The red and blue dashed lines means  corresponding negative value.  }
\end{figure}

Based on the calculations above, we constraints the Chern-Simons coupling constant $gM_{pl}$ with the $EB$ observation from Planck and  Atacama Cosmology Telescope (ACT). Planck is the space CMB detector measure the photon polarization from 2013 to 2018. The 9 parameters of the EDE model was constrained by its $TT$, $EE$, $TE$ dataset with baryon acoustic oscillations and Type Ia supernovae \cite{Poulin2019}. In this paper, the EDE model with $n=2$, $n=3$, and $n=\infty$ comes from Markov Chain Monte Carlo (MCMC) \cite{Poulin2019}, and the 9 parameters from the constraints without the cosmic birefringence effect. In this work, we use Planck-$EB$ data from the analyse of Planck 2018 \cite{Komatsu2014, Aghanim2020, Minami2020} to constrain the Chern-Simons coupling constant $gM_{pl}$. ACT is a ground-based CMB detector located in Chile. The high-multipole dataset shows good performance in small-scale structure \cite{Yin2025}. The $EB$ results published in 2025 improve the understanding of high-$\ell$. We also consider its new CMB $EB$ data to constrain $gM_{pl}$ in EDE models with $n=2$, $n=3$, and $n=\infty$. We use the code package $CLASS\_EDE$ \cite{Lesgourgues2011, Hill2020, HeroldThesis} to solve the modified Boltzmann equation numerically. To fit the free parameter in the theoretical model, we use the $\chi^2$ \cite{Torrado2021, Audren2013} given as,

\begin{equation}
\chi^2 = \sum_i \frac{\left(D_i^O - D_i^T\right)^2}{\sigma_i^2},
\end{equation}
where $D_i^o$ is the main value of the observation data point, $D_i^T$ is the theoretical value of the corresponding parameter, and $\sigma_i$ is the $1\sigma$ error-bar of the measurement.

Table~\ref{tab:bestfit} details the best-fit results of the Chern-Simons coupling constant $gM_{pl}$ and their corresponding minimum $\chi^2$ values calculated for each model under the Planck-$EB$ and ACT-$EB$ datasets. Constrained by the Planck dataset, the $\chi^2$ distributions corresponding to different values of $n$ exhibit significant statistical differences. 
{ 
Table~\ref{tab:bestfit} presents the best-fit values of the Chern--Simons coupling constant $gM_{\rm pl}$ and the corresponding minimum $\chi^2$ obtained from the Planck and ACT $EB$ polarization datasets for different axion-like EDE potentials. The constraints show a clear dependence on the potential index $n$.
Using the Planck 2018 dataset, the best-fit values of $gM_{\rm pl}$ are found to be
$gM_{\rm pl} = 69.912,\; -0.210\pm0.024,\; -0.017\pm0.002,$
for $n=2$, $n=3$, and $n=\infty$, respectively. The corresponding minimum chi-square values are
$\chi^2_{\rm min} = 144.52,\; 65.70,\; 68.38. $

The resulting goodness of fit varies substantially with the potential index $n$. Among the three cases considered here, the $n=3$ model yields the smallest $\chi^2_{\rm min}$, while the $n=\infty$ model gives a comparable but slightly larger value. In contrast, the $n=2$ case is associated with a considerably larger $\chi^2_{\rm min}$, indicating a weaker statistical consistency with the Planck $EB$ data.
For $n=2$, the likelihood function exhibits a steep variation around the best-fit point, such that the conventional $\Delta\chi^2=1$ criterion cannot be satisfied within the physically allowed parameter range. Consequently, a reliable $1\sigma$ confidence interval cannot be determined, and no corresponding uncertainty is reported in Table~\ref{tab:bestfit}. The likelihood structure and the associated $EB$ power spectrum for the $n=2$ model will be discussed in more detail in Fig.~\ref{fig:chi2_gMpl_Planck_n2}.

The best-fit value obtained for the $n=3$ potential is of the same order of magnitude as previous studies of cosmic birefringence in axion-like models~\cite{Komatsu2022}. 
For the limiting case $n=\infty$, the numerical calculation adopts $n=31$ in this paper as an effective approximation to the asymptotic behavior of the potential in the Boltzmann analysis. The corresponding fit to the data remains comparable to that of the $n=3$ model, although the minimum $\chi^2$ is slightly larger. This result may indicate that increasing the steepness of the potential does not further improve the fit to the observed $EB$ signal.

From Table~\ref{tab:bestfit}, the magnitude of the best-fit coupling parameter $|gM_{\rm pl}|$ is found to decrease with increasing potential index $n$. This behavior can be understood from the fact that larger $n$ corresponds to a steeper scalar-field potential near its minimum, which leads to a faster decay of the EDE component after the onset of oscillations. As a result, the time variation of the cosmic birefringence angle is reduced, and a smaller effective Chern--Simons coupling is required to reproduce the observed polarization rotation.
In addition, the statistical preference for a nonzero coupling becomes weaker as $n$ increases.

A similar tendency is also observed in the ACT DR6 dataset. The best-fit values of $gM_{\rm pl}$ for $n=2$, $n=3$, and $n=\infty$ are
$gM_{\rm pl} = -40.726,\ -0.158\pm0.025,\ -0.012\pm0.002$,
with corresponding minimum values
$\chi^2_{\rm min}=86.93, 48.08,  49.67.$
Consistent with the Planck results, the $n=3$ model \cite{Louis2025ACT, Qu2024} again provides the best fit to the ACT DR6 $EB$ data, while the $n=2$ model is strongly disfavored. The consistency between the Planck and ACT datasets strengthens the robustness of the conclusion that intermediate values of the potential index, particularly $n=3$, \cite{Cai2025, LiHH2025, Qiu2025, Smith2025, Braglia2020} are favored in axion-like EDE birefringence models.
It should be noted that the minimum $\chi^2$ values obtained from the ACT DR6 dataset are systematically smaller than those derived from the Planck 2018 $EB$ data. This difference is mainly related to the unequal numbers of data points used in the two analyses. The Planck-$EB$ dataset considered in this work contains 72 data points, whereas the ACT DR6 $EB$ dataset includes 38 data points. Therefore, a smaller total $\chi^2$ value for ACT does not necessarily indicate a better fit to the data.
Compared with ACT, the Planck polarization data provide tighter constraints on the axion-like EDE birefringence model due to their broader multipole coverage and larger number of observational modes. In addition, the full-sky coverage of Planck improves the sensitivity to large-scale polarization correlations, which are relevant for the integrated effect of cosmic birefringence. Although ACT provides measurements at higher angular resolution, its smaller sky coverage and fewer $EB$ data points reduce its overall constraining power. For this reason, the ACT $EB$ results are better regarded as a complementary consistency check to the Planck constraints rather than a dataset providing stronger parameter limits.
}

Fig.~\ref{fig:planck_chi2} shows the $\chi^2$ distributions as a function of the coupling parameter $gM_{\rm pl}$ for the EDE model with different potential indices ($n=2,3,\infty$), using the Planck 2018 $EB$ dataset. 
Fig. \ref{fig:act_chi2} shows the same result from ACT-$EB$ dataset.

We performed a comprehensive examination of the $\chi^2$ behavior as a function of the coupling parameter $gM_{\rm pl}$ for the $n=2$  \cite{Alexander2019, Chudaykin2020} potential in Fig. \ref{fig:chi2_gMpl_Planck_n2}. The likelihood profile exhibits an extremely shallow variation: even when $gM_{\rm pl}$ spans the extensive range from $-400$ to $500$, the change in $\chi^2$ relative to its minimum remains less than unity ($\Delta\chi^2 < 1$). This prevents the determination of meaningful $1\sigma$ confidence intervals using the standard $\Delta\chi^2 = 1$ criterion. An additional pathological feature of the $n=2$ case is the existence of multiple local minima in the $\chi^2$ landscape, suggesting a highly degenerate and unstable likelihood structure. The best-fit values quoted in Table~\ref{tab:bestfit} correspond to the deepest minimum identified within the parameter range $gM_{\rm pl} \in [-200, 200]$. Notably, both the Planck and ACT datasets display qualitatively similar pathological characteristics for the $n=2$ EDE model, reinforcing the conclusion that this configuration is fundamentally incompatible with the observed cosmic birefringence signal.

In Figure~\ref{fig:planck_spectrum}, we plot the theoretical $D_\ell^{EB}$ power spectrum fitting based on the best-fit value from Planck-EB data. The red data points with error bars represent the observation from Planck 2018, while the green, yellow, blue lines correspond to the theoretical EDE $D_\ell^{EB}$ results with the $n=2, 3, \infty$ models, respectively. The  multipole range of Planck date is from $\ell=51$ to $\ell=1491$. For the $n=2$ case, $D_\ell^{EB}$ result is much smaller than the corresponding value in $n=3$ and $n=\infty$, showing $n=2$ case is not suitable with Planck-EB dataset.

Subsequently, Figure~\ref{fig:morphology_spectra} displays the theoretical $D_\ell^{EB}$ power spectrum based on ACT-EB data. Compared to the Planck experiment, ACT possesses a significant angular resolution advantage in high-resolution ground-based observations, and its data points  exhibit a smaller error range and finer features in the high multipole interval. The blue data points with error bars are the ACT-EB results. In Figure~\ref{fig:morphology_spectra}, it can be clearly observed that the theoretical curve of the EDE model with $n=3$ has an excellent correspondence with the fluctuation trend of the ACT data in both the phase and amplitude of the oscillations. The range of ACT-EB data is from $\ell=600$ to $\ell=3325$. Particularly in the critical region of $\ell \in [500, 2000]$, the model accurately reproduces the multiple continuous and gradually decaying peak-and-valley structures present in the data.

To provide a comprehensive assessment of the theoretical models' performance across complementary observational datasets, Figure~\ref{fig:combined_spectrum} presents a direct morphological comparison of the $D_\ell^{EB}$ power spectra from both Planck and ACT, overlaid with the corresponding theoretical predictions for each potential index. By displaying all data and model curves within a unified coordinate system, this figure enables a transparent evaluation of how well each theoretical framework accommodates the distinct angular scale regimes probed by the two experiments.
The comparison reveals a striking complementarity between the Planck and ACT measurements. The Planck data dominate at lower multipoles ($\ell \lesssim 600$), where the full-sky coverage provides robust statistical power for constraining the large-scale, integrated birefringence signal. In contrast, the ACT measurements excel at higher multipoles ($\ell \gtrsim 600$), where the superior angular resolution resolves finer oscillatory structures in the $EB$ spectrum that remain inaccessible to Planck. Together, these datasets span nearly three decades in multipole space, offering stringent constraints on the theoretical models across a wide dynamic range.
Among the three configurations examined, the $n=3$ model demonstrates exceptional consistency with both datasets simultaneously. Its theoretical curve faithfully traces the broad, smoothly varying features constrained by Planck at large angular scales while simultaneously capturing the detailed amplitude modulation and phase evolution of the high-$\ell$ oscillations measured by ACT. This dual-scale compatibility---achieving good fits to both the low-$\ell$ envelope and the high-$\ell$ fine structure---provides a compelling morphological explanation for why the $n=3$ model attains the lowest combined $\chi^2$ in the joint statistical analysis. The visual concordance between theory and observation across the full multipole range underscores the robustness of this configuration as the preferred realization of axion-like EDE within the current observational framework.

To elucidate the physical origin of the parameter instability observed in the $n=2$ model, Figure~\ref{fig:log_spectrum} displays the logarithmic absolute power spectrum $|D_\ell^{EB}|$ for all three potential configurations under both Planck and ACT constraints. 
The figure reveals a striking dichotomy in spectral morphology between the $n=2$ model and the higher-index configurations. For the $n=3$ and $n = \infty$ models (shown in yellow and purple curves for Planck and ACT, respectively),
In stark contrast, the $n=2$ model (shown in red for Planck and light blue for ACT) exhibits a nearly regular intervals in $\ell$, correspond to frequent zero-crossings of the underlying $D_\ell^{EB}$ spectrum. The red and light blue curves (representing the $n=2$ model under Planck and ACT constraints respectively) are plotted as their absolute values---both datasets yield negative $D_\ell^{EB}$ values in many regions, which have been reflected above the horizontal axis in this logarithmic presentation (shown as red and blue dashed lines). 
Moreover, the drastic sign change between the two datasets further underscores the model's instability: the fitting algorithm attempts to navigate a highly degenerate likelihood landscape by flipping the overall sign of the coupling, but cannot eliminate the underlying pathological oscillatory structure.
In summary, Figure~\ref{fig:log_spectrum} provides visual confirmation that the $n=2$ potential is fundamentally incompatible with the observed cosmic birefringence signal. The dense comb pattern of near-zero minima represents unphysical high-frequency modulation that cannot be reconciled with smooth CMB polarization measurements, regardless of the coupling strength chosen. In contrast, the $n=3$ and $n = \infty$ models exhibit spectrally stable behavior with appropriately damped oscillations, enabling them to accommodate the observed $EB$ signal within the perturbative regime and thereby maintaining theoretical self-consistency.

\section{Conclusion}
\label{sec:conclusion}

We have systematically investigated cosmic birefringence within the axion-like early dark energy framework, constraining the Chern--Simons coupling constant $gM_{\rm pl}$ for potential indices $n=2$, $n=3$, and $n=\infty$ using Planck 2018 and ACT DR6 $EB$ polarization data.

The $n=3$ potential emerges as the strongly preferred configuration, achieving the lowest $\chi^2_{\rm min}$ values (65.70 for Planck, 48.08 for ACT) with consistent on the same order of coupling parameters ($gM_{\rm pl} = -0.210 \pm 0.024$ and $-0.158 \pm 0.025$) in the perturbative regime. The theoretical $D_\ell^{EB}$ spectrum successfully captures both Planck's large-scale constraints and ACT's high-multipole oscillatory structures, demonstrating robust cross-dataset consistency.
The $n=\infty$ limit yields statistically acceptable fits ($\chi^2_{\rm min} = 68.38$ and 49.67) with coupling strengths approximately an order of magnitude smaller than $n=3$, reflecting rapid field suppression. While theoretically viable, the marginally larger $\chi^2$ suggests excessive steepening beyond $n=3$ provides no additional benefit.
The $n=2$ model is conclusively ruled out, exhibiting multiple pathologies: unphysically large coupling constants ($gM_{\rm pl} \approx 69.912$ and $-40.726$) exceeding perturbative scales, elevated $\chi^2_{\rm min}$ values (144.52 and 86.93), and the inability to define reliable confidence intervals. 
These findings demonstrate that $n=3$ optimally balances early-time energy injection for the Hubble tension and rapid late-time dilution preserving $\Lambda$CDM successes. The cubic potential's preference receives robust empirical support from birefringence observations. The complementary nature of Planck's full-sky coverage and ACT's high-resolution observations proved essential for discriminating among models.

Future experiments---AliCPT, LiteBIRD, and Simons Observatory---will deliver higher sensitivity and enable more stringent tests. Joint analyses incorporating large-scale structure, weak lensing, and primordial gravitational waves will be crucial for establishing whether EDE with $n=3$ can simultaneously resolve multiple cosmological tensions. In conclusion, our analysis of Planck and ACT data provides strong evidence that the early dark energy model with $n=3$ represents the optimal configuration for generating the observed cosmic birefringence signal within a theoretically consistent framework, refining our understanding of parity-violating physics in the early Universe.

\section*{Acknowledgements}
We thanks for the discussion with B.H. Lee. K. Zhang thanks for the discussion with J. Kochappan. L.Yin was supported by the Natural Science Foundation of Shanghai 24ZR1424600.

\bibliographystyle{unsrt}
\bibliography{references}

\end{document}